\documentclass[conference]{IEEEtran}
\IEEEoverridecommandlockouts

 \usepackage[english]{babel}

\usepackage{amscd,amssymb}
\usepackage{amsfonts,amsmath,array}
\usepackage{graphicx}
\usepackage{longtable,cite,soul,wrapfig}

\usepackage[hidelinks]{hyperref}
\usepackage{multirow}
\usepackage{booktabs,colortbl}

\newtheorem{theorem}{\bf Theorem}[section]

\ifCLASSINFOpdf
\else
\fi
\begin{document}
%
% paper title
% can use linebreaks \\ within to get better formatting as desired
\title{Application of Volterra Equations to Solve Unit Commitment Problem of Optimised Energy\\ Storage \& Generation\thanks{This work is funded by the International science and technology
cooperation program of China under Grant No. 2015DFR70850.}}

% author names and affiliations
% use a multiple column layout for up to three different
% affiliations
\author{\IEEEauthorblockN{Ildar Muftahov\\ Denis Sidorov\\Aleksei Zhukov\\Daniil Panasetsky}
\IEEEauthorblockA{Energy Systems Institute\\Russian Academy of Sciences\\
Irkutsk, Russia \\
Email: dsidorov@isem.irk.ru}
\and
\IEEEauthorblockN{Aoife Foley }
\IEEEauthorblockA{
Sch. of Mechanical and Aerospace Engrg\\ Queen's University Belfast \\
Belfast, UK \\
Email: 	a.foley@qub.ac.uk}
\and
\IEEEauthorblockN{Yong Li\footnote{}}
\IEEEauthorblockA{Electrical Engrg Dept. \\ Hunan University \\
Changsha, China \\
Email: yongli@hnu.edu.cn}
\and
\IEEEauthorblockN{Aleksandr Tynda}
\IEEEauthorblockA{Computer Engrg  Dept.\\ Penza State University \\
Penza, Russia \\
\,\,Email: tynda@pnzgu.ru}
}
% conference papers do not typically use \thanks and this command
% is locked out in conference mode. If really needed, such as for
% the acknowledgment of grants, issue a \IEEEoverridecommandlockouts
% after \documentclass

% for over three affiliations, or if they all won't fit within the width
% of the page, use this alternative format:
% 
%\author{\IEEEauthorblockN{Michael Shell\IEEEauthorrefmark{1},
%Homer Simpson\IEEEauthorrefmark{2},
%James Kirk\IEEEauthorrefmark{3}, 
%Montgomery Scott\IEEEauthorrefmark{3} and
%Eldon Tyrell\IEEEauthorrefmark{4}}
%\IEEEauthorblockA{\IEEEauthorrefmark{1}School of Electrical and Computer Engineering\\
%Georgia Institute of Technology,
%Atlanta, Georgia 30332--0250\\ Email: see http://www.michaelshell.org/contact.html}
%\IEEEauthorblockA{\IEEEauthorrefmark{2}Twentieth Century Fox, Springfield, USA\\
%Email: homer@thesimpsons.com}
%\IEEEauthorblockA{\IEEEauthorrefmark{3}Starfleet Academy, San Francisco, California 96678-2391\\
%Telephone: (800) 555--1212, Fax: (888) 555--1212}
%\IEEEauthorblockA{\IEEEauthorrefmark{4}Tyrell Inc., 123 Replicant Street, Los Angeles, 

% use for special paper notices
%\IEEEspecialpapernotice{(Invited Paper)}

% make the title area
\maketitle

\begin{abstract}
%\boldmath
Development of reliable methods for optimised energy storage and generation is one of the most imminent challenges in modern power systems. In this paper an adaptive approach to load leveling problem using novel dynamic models based on the Volterra integral equations of the first kind with piecewise continuous 
kernels is proposed. These integral equations efficiently solve such inverse problem taking into account both  the time dependent efficiencies and the availability of generation/storage of each energy storage technology.
%of the different storages technologies and their proportions  in the total power generation. 
In this analysis a direct numerical method is employed to find the least-cost dispatch of available storages. The proposed collocation type numerical method has second order accuracy and enjoys self-regularization properties, which is associated with confidence levels of system demand.   This adaptive approach is suitable for energy storage optimisation in real time. The efficiency of the proposed methodology is demonstrated on the Single Electricity Markets of Republic of Ireland and Sakhalin island in Russian Far East.
\end{abstract}
% IEEEtran.cls defaults to using nonbold math in the Abstract.
% This preserves the distinction between vectors and scalars. However,
% if the conference you are submitting to favors bold math in the abstract,
% then you can use LaTeX's standard command \boldmath at the very start
% of the abstract to achieve this. Many IEEE journals/conferences frown on
% math in the abstract anyway.
% no keywords
% For peer review papers, you can put extra information on the cover
% page as needed:
% \ifCLASSOPTIONpeerreview
% \begin{center} \bfseries EDICS Category: 3-BBND \end{center}
% \fi
%
% For peerreview papers, this IEEEtran command inserts a page break and
% creates the second title. It will be ignored for other modes.
\IEEEpeerreviewmaketitle
\section*{Introduction}
% no \IEEEPARstart

%Future increases electrification, renewable energy usage and decentralization of transport and heating loads in power systems will result in an even more complex unit commitment problem (UCP). This will be further complicated in many countries by growth in variable renewable generation from wind, solar and even wave. 

Further growth in renewable energy and the planned electrification and decentralization of transport and heating loads in future power systems will result in a more complex unit commitment problem (UCP).

The traditional least cost dispatch of available generation  based on partial differential equations to meet load will no longer be fit for purposes for number of reasons including further decentralization of power systems (including transport and heating loads) and renewable energy usage.   A number of alternative methods have been examined to solve this challenging problem. For example Lagrangian relaxation, Mixed Integer Linear programming, particle swarm optimization, evolutionary methods and genetic algorithms, \cite{vse2013}. These methods have many advantages and disadvantages over the traditional approach. However, they have not been applied by industry due to the complicated heuristics and computational constraints. 
%The increasing penetration of renewable energy has led to increasing dependency on the volatile nature of renewable energy sources making UCP even more complicated.
The increasing complexity of the UCP can be seen in power systems with large wind penetrations.
% Several European countries have reached already a high level of installed wind power capacity, such as Germany, Spain, Denmark and Ireland, while others follow with fast rates of development. 
Several European countries, such as Germany, Spain, Denmark and Ireland, have already reached a high level of installed wind power, while others such as China and the USA are showing fast rates of development.
For example, wind farms in Eastern Germany during strong wind conditions can supply up to 12 GW, which is more than all of the coal- and gas-fired power plants in that part of the country combined. The Irish transmission system operator (TSO), currently restricts the instantaneous proportion of total generation allowed from non-synchronous sources, such as wind turbines, to 50 {\%} maximum, in order to maintain sufficient system inertia \cite{web}. 
This can result in wind curtailment at any time, for example during high wind speeds and low electricity demand or during sudden high speeds (i.e. ramps) during periods of high electricity demand, to maintain system stability.
This has an economic impact on the power system with increased operational costs through additional grid balancing charges.
Thus large-scale integration of wind power is challenging in terms of power system management.
This increase in the overall cost of the produced energy  limits the benefits of using renewable energy resources. 

A way of reducing the uncertainty associated with wind power production is to use forecasting tools. Wind power and load forecasting, over lead times of up to 48 hours, are useful to the market operator for creating 
day-ahead unit commitment and economic dispatch schedules. Many TSO also uses shorter-term wind forecasts to draw upon system reserves for short term balancing. Increasing the value of wind generation through the improvement of prediction systems performance is recognised as one of the priorities in wind energy research needs for the coming years.  In fact, a reduction of the forecast errors by a fraction of
a percent can lead to substantial increases in trading profits. For example, according
to \cite{bun,sor}, an increase of only 1\%
in predicting forecast error  caused an increase of 10 million pounds in operating
costs per year for one electric utility in the UK.

UCP for optimised energy storage and generation attracted many researchers during the last decade.  It involves solution of  of various problems including energy loss minimization, load leveling (peak load shaving/shifting), load forecasting and overall energy infrastructures management. 
In \cite{Kalkhambkar2016} an optimal placement methodology of energy storage is designed to improve energy loss minimization through peak shaving in the presence of renewable distributed generation. Agent-based distributed control scheme  for real time peak power shaving is proposed in  \cite{Sharma2016134}.   Price-based control system in conjunction with energy storage is analysed in \cite{Barzin2015505} for two applications: space heating in buildings and domestic freezers. It is shown that savings of up to 62.64\% per day can be achieved based on New Zealand electricity rates.
The review of China coal-fired power units peak regulation with a detailed presentation of the installed capacity, peak shaving operation modes and support policies is given in \cite{Gu2016723}.
 \cite{Zakeri2015569} cover the life cycle cost analysis of
various  energy storage technologies such as  pumped hydropower storage, compressed air energy storage, flywheel, electrochemical batteries supercapacitors, and hydrogen energy storage. Analysis of  time-of-use energy cost management using distributed electrochemical storages and optimum community energy storage system are correspondingly given in \cite{Graditi2016515} and \cite{Parra2016130}. 
An optimised model of hybrid battery energy storage system based on cooperative game model is proposed in \cite{Han2015643}.
 In \cite{Pazouki2016219} effectiveness of wind turbine, energy storage, and demand response programs in the deterministic and stochastic circumstances and influence of uncertainties of the wind, price, and demand are assessed on the  Energy Hub planning.
It is to be noted that UCP solver is integral part of Energy Hub which manage interconnection of heterogeneous energy infrastructures including renewable energy resources.

\subsection{Volterra models}

In this paper a methodology is developed to solve the UCP using systems of Volterra integral equations (VIE) of the first kind considering future power systems with battery energy storages of various efficiencies using load forecasts over lead times of 24 hours.
For bibliography of the state-of-art short-term forecasting methods readers may refer to
In \cite{ijai2015, mono}  the state-of-art of short term electric load forecasting is discussed.  
Power load forecasting can be carried out using many different techniques such as classical statistical methods of time series analysis, regressive models or methods based on artificial intelligence.
The state-of-the-art machine learning algorithms includes Random Forest (RF) \cite{breiman2001random}, Gradient Trees Boosting  (GTB) \cite{friedman2001greedy} and SVM with radial kernel (SVM RK) \cite{smola1997support}. 
In this case time series like approach was applied i.e. it is assumed that there are dependencies between recent obtained values and future load values.

Volterra equations naturally take into account the evolutionary character of  power systems. The kernels of employed equations have jump discontinuities along the continuous curves which starts at the origin. Such piecewise continuous kernels $K(t,s)$ take into account both efficiencies of the different storing technologies and their proportions (which could be time dependent) in the total power generation. 
Efficiencies of the different storages \cite{Zakeri2015569} may depend not only on their age and usage duration (correspondinly defined by variables $t$ and $s$), but also on state of charge. 
This can be examined using nonlinear VIE.

Studies of  linear Volterra 
integral equations  of the first kind with piecewise continuous kernels \eqref{e2015-01} have been initiated
in \cite{StudiaInformatica,Sidorov2013} and pursued in articles \cite{volt_systems,mar_sid_ISU, tyndasidmuft,Sidorov2014,Markova2014} followed by monograph \cite{mono}. 
%The solution
%of linear integral equations of the first kind is of course classical problem and has been addressed
%by numerous authors. But only few authors studied these equations in case of jump
%discontinuous kernels. 
In general,
VIE of the first kind can be solved by reduction to equations of the second kind,  regularization algorithms developed for Fredholm equations can be also applied as well as direct discretization methods.
On the other hand, it is known that solutions of integral equations of the first kind can be unstable and this is a well known  ill-posed problem. This is due to the fact that the Volterra operator maps the considered solution space into its narrow part only. Therefore, the inverse operator is not bounded and it is necessary to assess the proximity of the solutions and the proximity of the
%right parts
right-hand side
using the different metrics: the proximity of the
%right parts
right-hand side
should be in a stronger metric.
%����� ����, ��� �������� � \cite{Sidorov-book}, ������� ��������� ���� \eqref{e2015-03} ����� ��������� ������������ ���������� � ���� ��������������� ��� \(t\to 0\).
Moreover, as shown in \cite{mono}, solutions of the VIE can contain arbitrary constants and can be unbounded as \(t\to 0\). The problems of existence, uniqueness and asymptotic behavior of solutions of VIE are explored in \cite{StudiaInformatica, Sidorov2014, Sidorov2013}.

%Evolutionary integral equations are in the core of many mathematical models in
%physics, energetics, economics and ecology.
%Excellent historical overview of the results concerning the
%VIEs of the first kind is given by H. Brunner  in the paper ``1896 -- 1996:
%One hundred years of Volterra integral equations of the first kind'', also ref. to the bibliography in  monograph \cite{irl15}.
The theory of integral models of evolving systems was initiated  in the  works of  L. Kantorovich \cite{kantor73}, R. Solow \cite{solow}  and V. Glushkov \cite{glush} in the mid-20th Century. 
It is well known that Solow publication \cite{solow} on vintage capital model led him to the Nobel prize in 1987 for his analysis of economic growth.

Such theory leads the VIEs of the first kind where bounds of the integration interval can be  functions of time. 
These models take into account
the memory of a dynamical system when its past impacts its future evolution.
The memory is implemented in the existing technological and financial structure of physical capital (equipment).
The memory duration is determined by the age of the oldest capital unit (e.g. storages) still employed.
There are several approaches available for numeric solution of VIE of the first kind. One of them is to apply classical regularizing algorithms developed for Fredholm integral equations of the first kind \cite{Kythe}. However, the problem reduces to solving algebraic systems of equations with a full matrix, an important advantage of the VIE is lost and there is a significant increase in arithmetic complexity of the algorithms. The second approach is based on a direct discretization of the initial equations. Here one may face an instability of the approximate solution %to
because of errors in the initial data (load forecast). The regularization properties of the direct discretization methods are optimal in this sense, where the discretization step is the regularization parameter associated with the error of the source data (load forecast error upper bound). 
%However, only low-order quadrature formulas (midpoint quadrature or trapezoidal formulas) are suitable for approximation of the integrals. .
 The detailed description of regularizing direct numerical algorithms is given in  \cite{Kythe}.
It is to be noted that
it is very difficult to apply
these algorithms to solve the equation \eqref{e2015-01} in the form of \eqref{e2015-03} %is very difficult
because of %breaks of the kernels
the kernel discontinuities \eqref{e2015-02}.
%������  - ������ �������������, ���������� �� �������-���������� � �������-�������� ������������� ������� ������� (������� � ������� ������� ��������, ��������������).
The adaptive mesh should depend on the curves of the jump discontinuity for each number \(N\) of divisions of the considered interval and therefore this mesh can not be linked to the errors in the source data. It is needed to correctly approximate the integrals.

It should be noted that  the load leveling problem in UCP has no such luxury as discretization step control because source data is measured using the fixed step only. Therefore Lavrentiev regularization \cite{Kythe, muft_sid_ISU} using parameter $\alpha$ (Fig. 2) is employed.

\paragraph{Linear Volterra models}

%�������� ���������� ������������ � ������ ������ �������� ������������ ��������� I ���� ���������� ����
The object of  interest is the following integral equation of the first kind
\begin{equation}\label{e2015-01}
  \int_{0}^{t}K(t,s)x(s)\;ds=f(t), \; t\in[0,T],
\end{equation}
%��� ���� \(K(t,s)\) ������ �������� ������� �� ������ \(\alpha_i(t),\;i=1,2,\ldots,n-1,\) � ����� ���
where the kernel \(K(t,s)\) is discontinuous along continuous curves \(\alpha_i(t),\;i=1,2,\ldots,n-1,\) and is of the form
\begin{equation}\label{e2015-02}
  K(t,s)=\left\{
           \begin{array}{ll}
             K_1(t,s), &  \alpha_0(t)< s < \alpha_1(t); \\
             K_2(t,s), &  \alpha_1(t)< s < \alpha_2(t); \\
             \cdots  \\
             K_n(t,s), &  \alpha_{n-1}(t)<  s < \alpha_n(t). \\
           \end{array}
         \right.
\end{equation}
%�����
Here
$
 \alpha_0(t)\equiv 0,\; \alpha_0(t)<\alpha_1(t)<\ldots<\alpha_n(t)\equiv t,\; f(0)=0.
$
%�����������, ��� ���� \(K_i(t,s)\) � ������ ����� \(f(t)\) � ��������� \eqref{e2015-01} �������� ������������ � ���������� �������� ���������. ������� \(\alpha_i(t)\in C^1[0,T]\) � �������� ������������. ����� ����
Assuming the kernels \(K_i(t,s)\) and the
%right part
right-hand side
\(f(t)\) in the equation \eqref{e2015-01} are continuous and sufficiently smooth functions. The functions \(\alpha_i(t)\in {\mathcal C}^1[0,T]\) are not decrescent. Moreover
$
  \alpha_1'(0)\le\alpha_2'(0)\le\ldots\le\alpha_{n-1}'(0)<1.
$
%��������� ��������� \eqref{e2015-01} � ����������� ����
Rewriting the equation \eqref{e2015-01}
$$  \int_{0}^{\alpha_1(t)}K_1(t,s)x(s)\;ds+ \int_{\alpha_{1}(t)}^{\alpha_2(t)}K_2(t,s)x(s)\;ds+
  \cdots $$
\begin{equation}\label{e2015-03}
 \cdots+ \int_{\alpha_{n-1}(t)}^{t}K_n(t,s)x(s)\;ds=f(t), \; t\in[0,T].
\end{equation}
%It is to be noted that conventional Glushkov integral model of evolving systems is the special case of this equation  where all the functions $K_i(t,s)$
% are zeros except of $K_n(t,s)$. 
%For more details concerning evolving (developing) systems modeling using integral models with discontinuous kernels readers may refer to \cite{Sidorov-book}, \cite{ApSid}.

%************************************************************************************************************
%\section{������ �������������}
%\subsection{�������-���������� �������������}
\subsection{Numerical solution}

%\noindent{\it Piecewise constant approximation}\\

%\noindent{\it Piecewise linear approximation}\\

%��� ���������� ���������� ������� ��������� \eqref{e2015-03} �� ������� \([0,T]\) (� �������� ������������� ������������� ������������ �������) ������ ����� ����� (������������� �����������)
The mesh nodes are introduced (not necessarily uniform) to construct the numeric solution of the equation \eqref{e2015-03} on the interval \([0,T]\)
%(in conditions of existence of a unique continuous solution)
(if the unique continuous solution exists)
%\begin{equation}\label{e2015-04}
  $0=t_0<t_1<t_2<\ldots<t_N=T, $
%\end{equation}
$h=\max\limits_{i=\overline{1,N}}(t_i-t_{i-1})=  {\mathcal O}(N^{-1}).$
The approximate solution is determined as the following piecewise linear function
\begin{equation}\label{e2015-13}
   x_N(t)=\sum\limits_{i=1}^{N}\left(x_{i-1}+\frac{x_i-x_{i-1}}{t_i-t_{i-1}}(t-t_{i-1})\right)\delta_i(t),\; t\in(0,T], \;
\end{equation}
%���, ��� � �����,
where
$
   \delta_i(t)=\left\{
            \begin{array}{ll}
              1, & \hbox{for } t\in \Delta_i=(t_{i-1},t_i]. \\
              0, & \hbox{for } t\notin\Delta_i.
            \end{array}
          \right.
$
%������������ \(x_i,\;i=\overline{1,N},\) ������������� ������� �������� �����������.
%��������� �� ������� \eqref{e2015-06} ����������� \(x_0\)  � �������� ��������� \eqref{e2015-10}, �������
The objective is to determine the coefficients \(x_i,\;i=\overline{1,N},\) of the approximate solution.
Both parts of the equation \eqref{e2015-03} are differentiated with respect to \(t\) to determine \(x_0=x(0)\)\\
$f'(t)=$
\[
 \sum\limits_{i=1}^n \Biggl( \int_{\alpha_{i-1}(t)}^{\alpha_i(t)}
       \frac{\partial K_i(t,s)}{\partial t}x(s)\;ds+\alpha'_i(t)K_i(t,\alpha_i(t))x(\alpha_i(t))-
\]
\[
       -\alpha'_{i-1}(t)K_i(t,\alpha_{i-1}(t))x(\alpha_{i-1}(t))\Biggr).
\]
From the last expression coefficient $x_0$ is obtained:
\begin{equation}\label{e2015-06}
   x_0=\frac{f'(0)}{\sum\limits_{i=1}^n K_i(0,0)\left[\alpha'_i(0)-\alpha'_{i-1}(0)\right]}.
\end{equation}
It is assumed
%equations \eqref{e2015-03} are such
that the denominator of expression \eqref{e2015-06} is not zero.
%пїЅпїЅпїЅпїЅпїЅпїЅ пїЅпїЅпїЅпїЅпїЅ пїЅпїЅпїЅпїЅпїЅпїЅпїЅпїЅпїЅпїЅпїЅ \(f_k=f(t_k), \;k=1,\ldots,N\). пїЅпїЅпїЅ пїЅпїЅпїЅпїЅпїЅпїЅпїЅпїЅпїЅпїЅпїЅ пїЅпїЅпїЅпїЅпїЅпїЅпїЅпїЅпїЅпїЅпїЅпїЅ \(x_1\) пїЅпїЅпїЅпїЅпїЅпїЅпїЅ пїЅпїЅпїЅпїЅпїЅпїЅпїЅпїЅ пїЅпїЅпїЅпїЅпїЅпїЅпїЅпїЅпїЅ пїЅ пїЅпїЅпїЅпїЅпїЅ \(t=t_1\):
Make the notation \(f_k:=f(t_k), \;k=1,\ldots,N\) and write the initial equation in the point \(t=t_1\) to define the coefficient \(x_1\):
\begin{equation}\label{e2015-07}
  \sum\limits_{i=1}^n  \int_{\alpha_{i-1}(t_1)}^{\alpha_i(t_1)}K_i(t_1,s)x(s)\;ds=f_1.
\end{equation}
%пїЅпїЅпїЅ пїЅпїЅпїЅ пїЅпїЅ пїЅпїЅпїЅпїЅпїЅпїЅ пїЅпїЅпїЅпїЅ пїЅпїЅпїЅпїЅпїЅ пїЅпїЅпїЅпїЅ пїЅпїЅпїЅпїЅпїЅпїЅпїЅпїЅ пїЅпїЅпїЅпїЅпїЅпїЅпїЅпїЅпїЅпїЅпїЅпїЅпїЅпїЅ \(\alpha_i(t_1)-\alpha_{i-1}(t_1)\) пїЅ \eqref{e2015-07} пїЅпїЅ пїЅпїЅпїЅпїЅпїЅпїЅпїЅпїЅпїЅпїЅпїЅ \(h\), пїЅ пїЅпїЅпїЅпїЅпїЅпїЅпїЅпїЅпїЅпїЅпїЅпїЅ пїЅпїЅпїЅпїЅпїЅпїЅпїЅ пїЅпїЅпїЅпїЅпїЅпїЅпїЅпїЅпїЅ пїЅпїЅпїЅпїЅпїЅпїЅпїЅпїЅ \(x_1\), пїЅпїЅ, пїЅпїЅпїЅпїЅпїЅпїЅпїЅпїЅ пїЅпїЅпїЅпїЅпїЅпїЅпїЅпїЅпїЅпїЅпїЅпїЅ пїЅпїЅпїЅпїЅпїЅпїЅпїЅ пїЅпїЅпїЅпїЅпїЅпїЅпїЅ пїЅпїЅпїЅпїЅпїЅпїЅпїЅпїЅпїЅпїЅпїЅпїЅпїЅпїЅпїЅ, пїЅпїЅпїЅпїЅпїЅ
{Since at this stage the lengths of all integration intervals \(\alpha_i(t_1)-\alpha_{i-1}(t_1)\) in equation \eqref{e2015-07} do not exceed \(h\) then based on the midpoint quadrature, the result is equation \eqref{e2015-08}.
\begin{equation}\label{e2015-08}
   x_1=\frac{f_1}{\sum\limits_{i=1}^n (\alpha_i(t_1)-\alpha_{i-1}(t_1))K_i\left(t_1,\frac{\alpha_i(t_1)+\alpha_{i-1}(t_1)}{2}\right)}
\end{equation}
%пїЅпїЅпїЅпїЅпїЅпїЅпїЅпїЅпїЅпїЅпїЅ пїЅпїЅпїЅпїЅпїЅпїЅ, пїЅпїЅпїЅ пїЅпїЅпїЅ пїЅпїЅпїЅпїЅпїЅпїЅпїЅ пїЅпїЅпїЅпїЅпїЅпїЅпїЅпїЅ \(x_2,x_3,\ldots,x_{k-1}\).
%пїЅпїЅпїЅпїЅпїЅпїЅпїЅпїЅпїЅ пїЅпїЅпїЅпїЅпїЅпїЅпїЅпїЅпїЅ \eqref{e2015-01} пїЅ пїЅпїЅпїЅпїЅ
Suppose the values \(x_2,x_3,\ldots,x_{k-1}\) are known.
Then equation \eqref{e2015-01} can be written as follows
\begin{equation}\label{e2015-09}
   \int_{t_{k-1}}^{t}K(t,s)x(s)\;ds=f(t)- \int_{0}^{t_{k-1}}K(t,s)x_N(s)\;ds
\end{equation}
%пїЅ пїЅпїЅпїЅпїЅпїЅпїЅпїЅпїЅпїЅ пїЅпїЅпїЅпїЅпїЅпїЅпїЅпїЅпїЅпїЅ пїЅпїЅпїЅпїЅпїЅпїЅпїЅпїЅпїЅпїЅ пїЅпїЅпїЅпїЅпїЅпїЅпїЅпїЅпїЅ пїЅ пїЅпїЅпїЅпїЅпїЅ \(t=t_k\):
and  require that the last equality holds for the point \(t=t_k\)
\begin{equation}\label{e2015-10}
   \int_{t_{k-1}}^{t_k}K(t_k,s)x(s)\;ds=f_k- \int_{0}^{t_{k-1}}K(t_k,s)x_N(s)\;ds.
\end{equation}
The coefficient \(x_0\) is determined by expression \eqref{e2015-06} and taking into account the equality \eqref{e2015-10} 
\[
   \int_{t_{k-1}}^{t_k}\left(x_{k-1}+\frac{x_k-x_{k-1}}{t_k-t_{k-1}}(s-t_{k-1})\right)K(t_k,s)\;ds=
\]
\[
  =f_k-\sum\limits_{j=1}^{k-1} \int_{t_{j-1}}^{t_{j}}\left(x_{j-1}+\frac{x_j-x_{j-1}}{t_j-t_{j-1}}(s-t_{j-1})\right)K(t_k,s)\;ds.
\]
%������, �������� \(x_k\), �����
Thus excluding \(x_k\),
$$ x_k=x_{k-1}+ $$ 
$$
 +\biggl[f_k-x_{k-1} \int_{t_{k-1}}^{t_k}K(t_k,s)\;ds-
  \sum\limits_{j=1}^{k-1}\bigl(x_{j-1} \int_{t_{j-1}}^{t_j}K(t_k,s)\;ds+$$
$$+{x_j-x_{j-1}}{t_j-t_{j-1}} \int_{t_{j-1}}^{t_j}(s-t_{j-1})K(t_k,s)\;ds \bigr)\biggr]/$$ 
\begin{equation}\label{e2015-14}
 \biggl\{\frac{1}{t_k-t_{k-1}} \int_{t_{k-1}}^{t_k}(s-t_{k-1})K(t_k,s)\;ds\biggr\},\,
k=1,2,\ldots,N.
\end{equation}
 The integrals in \eqref{e2015-14} can be approximated using the midpoint quadratures based on auxiliary mesh nodes so that the values of the functions \(\alpha_i(t_j)\) are a subset of the set of this mesh points at each particular value of \(N\).
%����������� ������ ��� ����� ������������� ����� ���:
The error of this approximation method is
%\begin{equation}\label{e2015-15}
$ \varepsilon_N=\|x(t)-x_N(t)\|_{C_{[0,T]}}={\mathcal O}\left({1}/{N^2}\right).$
\paragraph{Nonlinear equations}
In this paragraph the following nonlinear equation is addressed
\begin{equation}\label{e2015-31}
   \int_0^t K(t,s,x(s))\;ds = f(t), \quad t\in[0,T], \;f(0)=0,
\end{equation}
where
\begin{equation}\label{e2015-32}
    K(t,s,x(s)) = \left\{ \begin{array}{ll}
         \mbox{$K_1(t,s)G_1(s,x(s)), \,\, t,s \in m_1$}, \\
         \mbox{\,\, \dots \,\, \dots \dots \dots } \\
         \mbox{$K_n(t,s)G_n(s,x(s)), \,\, s \in m_n$}. \\
        \end{array} \right.
\end{equation}
Here \( m_i = \{ t,s  \bigl | \alpha_{i-1}(t) <  s < \alpha_i(t) \},\)   \( \alpha_0(t)=0,\; \alpha_n(t)=t,\; i=\overline{1,n}\),
The functions  \(f(t)\),  \(\alpha_i(t)\) have continuous derivatives with respect to  $t$ in the corresponding domains ${m_i},$  $ K_n(t,t) \neq 0,$   $\alpha_i(0)=0,$ $\,\,\,\,0 < \alpha_1(t)<\alpha_2(t)< \cdots < \alpha_{n-1}(t)<t$.
Functions $K_i, i=1, \dots ,n$ have continuously differentiable w.r.t. $t$ continuation into compacts $\overline{m}_i$.
The functions $\alpha_1(t), \dots , \alpha_{n-1}(t) $  should increase at least in small neighbourhood of origin.
Proof of the existence and uniqueness theorem of equation \eqref{e2015-31} is similar with proof of the Theorem 3.2 in \cite{mono}. In \cite{SidBlowUp} it is outlined that existence theory for such nonlinear equations is chellanging problem even in case of continuous kernels. 

The nonlinear integral operator is introduced in order to approximate solutions for equation 
\eqref{e2015-31}.
\begin{equation}\label{e2015-33}
  (Fx)(t)\equiv \sum\limits_{i=1}^{n} \int_{\alpha_{i-1}(t)}^{\alpha_i(t)} K_i(t,s)G_i(s,x(s))\;ds - f(t)
\end{equation}
Equation \eqref{e2015-31} can be written in an operator form such that
\begin{equation}\label{e2015-34}
  (Fx)(t)=0
\end{equation}
The operator \eqref{e2015-33} can be linearized according to a
modified Newton-Kantorovich scheme \cite{Kantorovich2} In order to construct an iterative numerical method for equation \eqref{e2015-31}:
\begin{equation}\label{e2015-35}
     x_{m+1}=x_m-[F'(x_0)]^{-1}(F(x_m)) ,\; m=0,1,\ldots,
\end{equation}
where \(x_0(t)\) is the initial approximation. Then, the approximate solution of \eqref{e2015-34} could be determined as the following limit of sequence:
\begin{equation}\label{e2015-36}
  x(t)=\lim\limits_{m\to\infty}x_m(t).
\end{equation}
The derivative \(F'(x_0)\) of the nonlinear operator \(F\) at the point \(x_0\) is defined as follows:
\[
  F'(x_0)=\lim_{\omega\to 0}\frac{F(x_0+\omega x)-F(x_0)}{\omega}=
\]
\[
  =\lim_{\omega\to 0}\frac{1}{\omega}\sum\limits_{i=1}^{n}\int_{\alpha_{i-1}(t)}^{\alpha_i(t)}
  K_i(t,s)\bigl[G_i(s,x_0(s)+\omega x(s))-\] \[-G_i(s,x_0(s))\bigr]\;ds.
\]
Implementing the limit transition under the integral sign such that
\begin{equation}\label{e2015-37}
    F'(x_0)(t)=\sum\limits_{i=1}^{n}\int_{\alpha_{i-1}(t)}^{\alpha_i(t)}
    K_i(t,s)G_{ix}(s,x_0(s))x(s)\;ds, 
\end{equation}
$\text{ where } G_{ix}(s,x_0(s))=\left.\frac{\partial G_i(s,x(s))}{\partial x}\right|_{x=x_0}.$
%%-----------------------
Thus,  the operator form of the Newton-Kantorovich scheme is obtained as follows:
\begin{equation}\label{e2015-38}
    F'(x_0(t))\Delta x_{m+1}(t)=-F(x_m), \; \Delta x_{m+1}=x_{m+1}-x_m,
\end{equation}
or in the extended form
$$
\sum\limits_{i=1}^{n}\int_{\alpha_{i-1}(t)}^{\alpha_i(t)}
    K_i(t,s)G_{ix}(s,x_0(s))\Delta x_{m+1}(s)\;ds= $$
\[
   =f(t)-\sum\limits_{i=1}^{n}\int_{\alpha_{i-1}(t)}^{\alpha_i(t)}
    K_i(t,s)G_{i}(s,x_m(s))\;ds.
\]
%%-----------------------
The latter can be written as follows
\begin{equation}\label{e2015-39}
    \sum_{i=1}^{n}\int_{\alpha_{i-1}(t)}^{\alpha_i(t)}
    K_i(t,s)G_{ix}(s,x_0(s)) x_{m+1}(s)\;ds=\Psi_m(t), 
\end{equation}
where
\[
  \Psi_m(t)=f(t)+\sum_{i=1}^{n}\int_{\alpha_{i-1}(t)}^{\alpha_i(t)}
   K_i(t,s)\times 
\]
$$\times \left[G_{ix}(s,x_0(s)) x_{m}(s)-G_{i}(s,x_m(s))\right] \,ds.$$
Equations \eqref{e2015-39} are now linear VIE with respect to the unknown function \(x_{m+1}(t)\).
Note that the kernels \(K_i(t,s)G_{ix}(s,x_0(s))\), \(i=\overline{1,n},\)   remain constant during each iteration \(m\).
Since equations \eqref{e2015-39} have the form \eqref{e2015-03} the method suggested in Section B can be applied.
Thus, solving the equations \eqref{e2015-39}  a sequence of approximate functions \(x_{m+1}(t)\) is achieved. Then using formula \eqref{e2015-36},
 the approximate solution of equation \eqref{e2015-31} with an accuracy depending on \(m\)
is achieved.

%\subsection{The convergence theorem }
Let \({\mathcal C}[0,T]\) be a Banach space of continuous functions equipped
with the standard norm \( \|x\|_{{\mathcal C}[0,T]}=\max\limits_{t\in [0,T]}|x(t)| \).
%The following theorem of convergence (based on the general theory proposed in the classical monograph
%\cite{Kantorovich2}) for iterative process \eqref{e2015-39} takes
%place:
%%------------------
\begin{theorem}
   Let the operator \(F\) have a continuous second derivative in the sphere
   \(\Omega_0\;(\|x-x_0\|\leqslant r)\) and the following conditions
   hold:
   \begin{enumerate}
      \item Equation \eqref{e2015-39} has a unique solution in
      \([0,T]\) for \(m=0\), i.e. there exists \(\Upsilon_0=[F'(x_0)]^{-1}\);
      \item \(\|\Delta x_1\|\leqslant\eta; \)
      \item \(\|\Upsilon_0F''(x)\|\leqslant L,\;\;x\in\Omega_0\).
   \end{enumerate}
If also
\(
  h=L\eta<\frac12 \text{ and }
  \frac{1-\sqrt{1-2h}}{h}\eta\leqslant
  r\leqslant\frac{1+\sqrt{1-2h}}{h}\eta,
\)
then equation \eqref{e2015-31} has a unique solution \(x^*\) in
\(\Omega_0\), process \eqref{e2015-39} converges to \(x^*\), and the
velocity of convergence is estimated by the inequality $
  \|x^*-x_m\|\leqslant \frac{\eta}{h}(1-\sqrt{1-2h})^{m+1},
  \;m=0,1,\ldots. $
\end{theorem}
In order to prove this theorem one must show that equation
\eqref{e2015-39} is uniquely solvable (including the case \(m=0\)),
i.e. condition 1 of the theorem holds. Then the
boundedness of the second derivative \(\bigl[F''(x_0)\bigr](x)\)) cab be verified
to estimate the constant \(L\) in cond. 3.

%One can verify that the necessary condition for the second
%derivative \(\bigl[F''(x_0)\bigr](x)\) to be bounded is a
%differentiability of the initial approximation \(x_0(t)\) as well as
%the functions \(K_i\) with respect to second variable.

%%------------------------------------------------------------------------------------
%%------------------------------------------------------------------------------------

\section{Results of Numerical Experiments on Real Data}

In order to evaluate proposed load leveling method the power load time series are taken  from Single Electricity Market of Republic of Ireland (SEM ROI) \cite{eirgrid} and Sakhalin island (SEM SI) in Russian Far East. The time interval is one week from 25.04.2016 till 01.05.2016. It is assumed that part of the batteries are eventually removed from service causing efficiency reduction up to 15\%. This phenomena is described below by coefficients of the Volterra kernel 
\eqref{voltker}.

\subsection{Load Forecasting}

RF, GTB and SVM RK are evaluated to obtain high accuracy model for power load forecast which is integral part of our approach. 
 In order to make 24 hours ahead forecast 6 recently received load values and day of the week were used as the most informative features.  Features importance was estimated based on RF Gini-importance measure \cite{breiman2001random}.
The models parameters were obtained using repeating cross-validation. RFs were used with $mtry = 5$ and 150 trees. In case of GTB 200 trees were employed with $interaction.depth = 10$, $shrinkage = 0.1$ and $minobsinnode = 10$. As for SVM, the parameters were selected as $sigma = 0.4428387$ and $C = 128$.
 The performance of the methods was assessed using the mean absolute error (MAE) and root mean square error (RMSE).  
%All exploited methods show similar errors as shown in Tab. \ref{tab:error}.
%The metrics are defined as follows:
%
%\begin{equation}
%    \label{eq:MAE}
%    MAE = \frac{1}{n}\sum_{t=1}^{n}|x_t-\bar{x_t}|
%\end{equation}    
%
%\begin{equation}
%    \label{eq:RMSE}
%    RMSE = \sqrt{\frac{1}{n}\sum_{t=1}^{n}(x_t-\bar{x_t})^2} 
%\end{equation}    

%\begin{table}[ht]
%    \centering
%   \caption{Forecasting errors for various models} 
%    \label{tab:error}
%    \begin{tabular}{|l|l|l|l|}
%        \hline
%        & RF & GTB & SVM (RK) \\ 
%        \hline
%        RMSE & 224.08 & \textbf{209.93} & 227.15 \\ 
%        MAE & 161.06 & \textbf{151.02} & 162.68 \\
%        \hline    
%    \end{tabular}
%\end{table}
Fig.~1 and Fig.~2 show actual load and its forecasts with 24 hrs lead interval for SEM ROI and SEM SI. 
Here the solid green curve shows base generation (without storage), the solid blue curve stands for  actual load, the green dotted line corresponds to RF based forecast,  SVM forecast marked with  dashed red, black dashed dots stands for GTB.
All exploited methods show similar errors.

\begin{figure*}[htbp]
\begin{center}
	\includegraphics[scale=0.41]{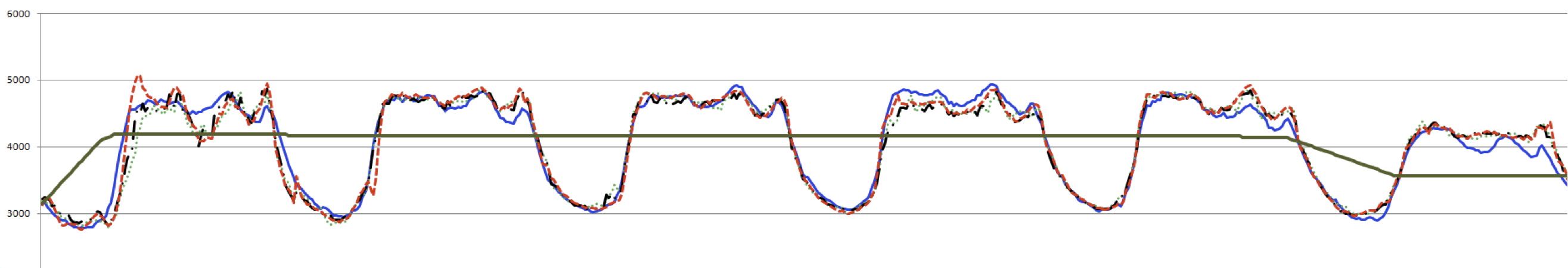}	
	\label{f1}
	\caption{Load forecasts with 24~hrs lead interval, actual load (blue curve) and available base generation (dark green line) for SEM ROI.}
\end{center}
\end{figure*}

\subsection{Load Leveling}

\begin{figure*}[htbp]
	\includegraphics[scale=0.556]{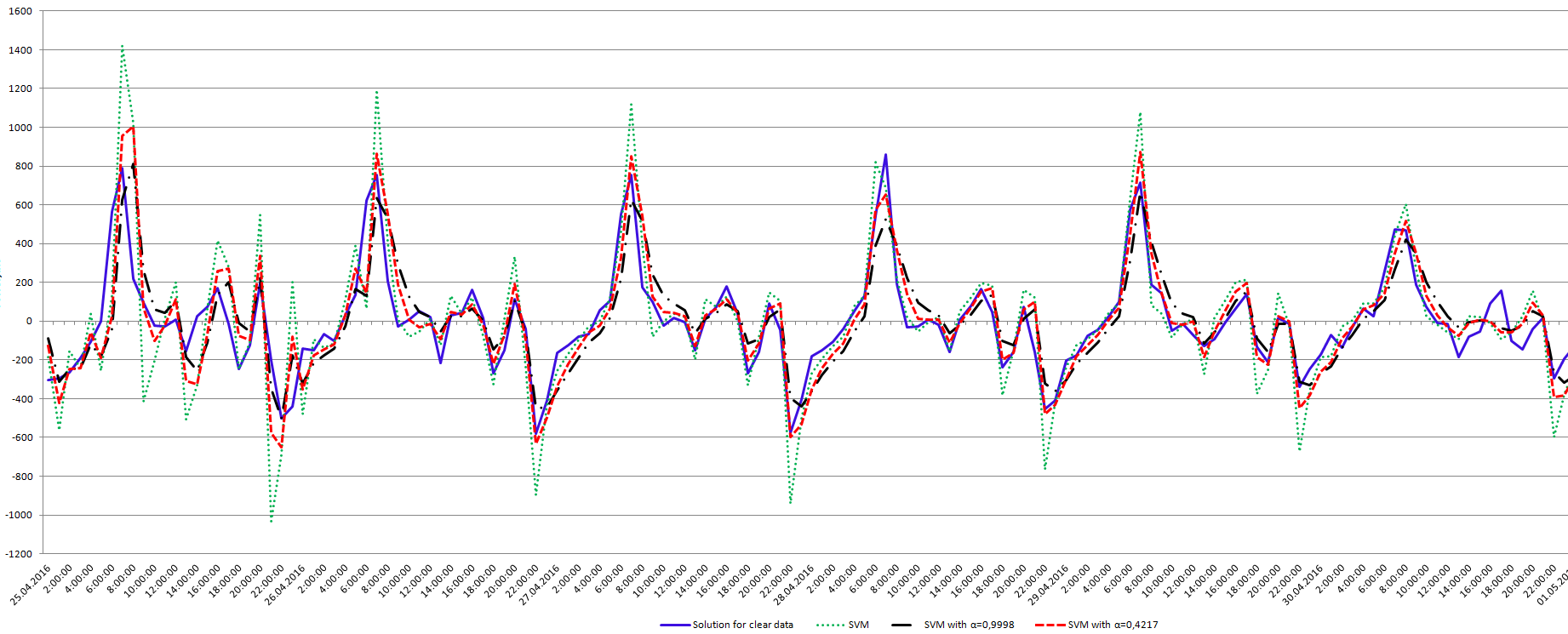}	
	\label{f2}
	\caption{Charge / discharge strategies using various 24 hrs forecasts comparing with actual data based benchmark strategy (sold line) for SEM ROI.}
\end{figure*}
%
%
%Unit commitment problem statement goes here in terms of systems of VIE
%Fig.~3 shows activity diagram of proposed approach. 
For sake of simplicity  the single storage usage is considered.  Assume its efficiency reduction during the test period (one week) and we  model this process
using VIE (1) with kernel
\begin{equation}
  K(t,s)=\left\{
           \begin{array}{ll}
             1, &  0< s < t/4 \\
              0.9, &  t/4< s < 3t/4\\
             0.85, &  3t/4< s < t\\
           \end{array}
         \right.\,\, t\in [0,\,T].
\label{voltker}
\end{equation}
Here coefficients 1, 0.9 and 0.85 corresponds to 100\%, 90\% and 85\% efficienties for the storage, see e.g. overview \cite{Zakeri2015569}.

\begin{table}[htb]%
\begin{center}
\caption{Charge/discharge errors for various $\alpha$ using SVM forecast for SEM ROI}\label{tab3_0_our-1}
\begin{tabular}{|c|c|c|c|}
	\hline
			&	SVM		&	SVM with $\alpha=0.4217$	&	SVM with $\alpha=0.9998$\\ \hline
	RMSE	&	194.6	&	142.48						&	145.52\\ \hline
	MAE		&	129.09	&	95.64						&	107.62\\ \hline
\end{tabular}
\end{center}
\end{table}

Fig.~2 and Fig.~4  shows charging/discharging strategies using various 24 hrs forecasts and actual data for SEM ROI and SEM SI correspondingly. Here the solid curve 
marks the strategy based on actual load, the dotted curve shows the strategy using a SVM forecast.
The dashed curve shows the strategy defined using SVM forecast and Lavrentiev regularization parameter $\alpha =0.4217$, the dashed dots curve is used to mark strategy calculated using the SVM forecast with~$\alpha =0.9998.$ 

As footnote, there interesting fact can be observed concerning charging/discharging strategies
comparison for Ireland and Sakhalin: there is strong single morning peak in SEM ROI, 
and there are two daily peaks in case of Sakhalin data.

\begin{figure*}[htbp]
\begin{center}
	\includegraphics[scale=0.63]{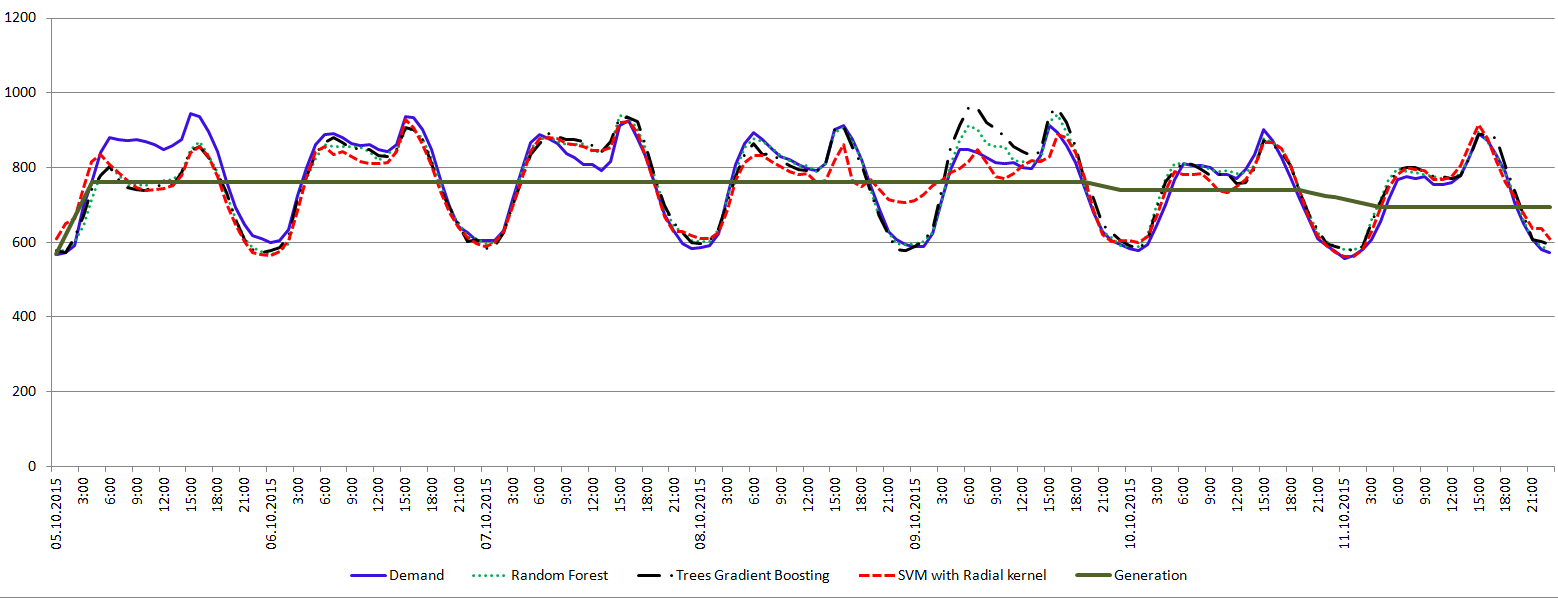}	
	\label{f3}
\end{center}
	\caption{Load forecasts with 24~hrs lead interval, actual load (blue curve) and available base generation (dark green line) for SEM SI.}
\end{figure*}

\begin{figure*}[htbp]
\begin{center}
	\includegraphics[scale=0.612]{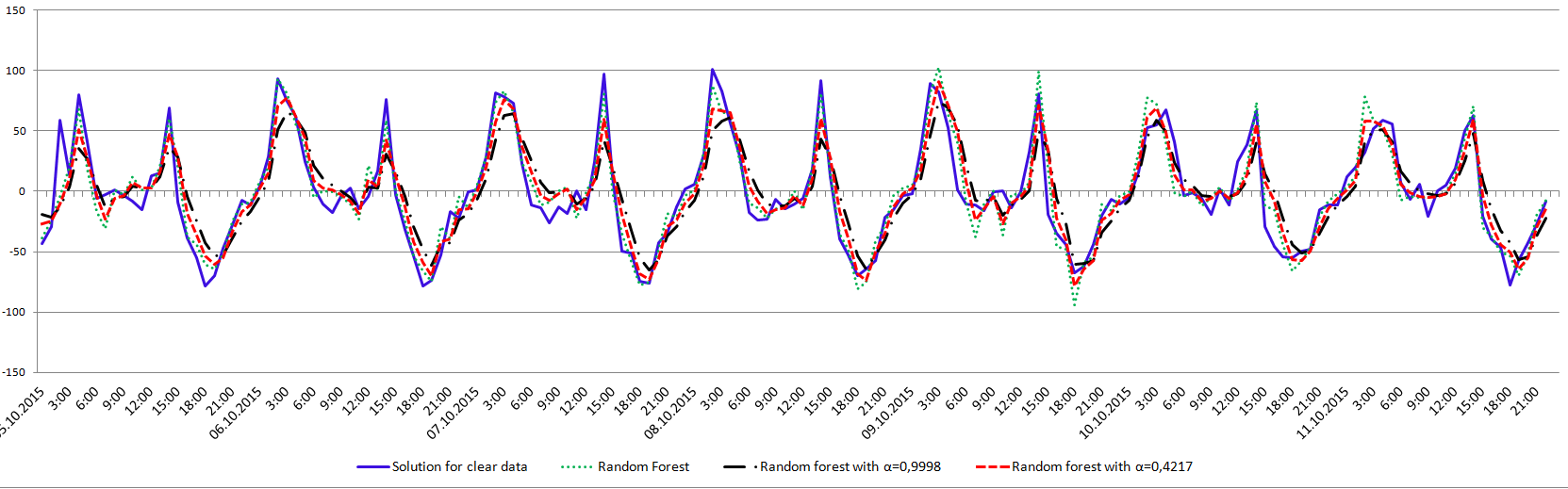}	
	\label{f4}
\end{center}
	\caption{Charge / discharge strategies using  24 hrs RF  forecasts comparing with actual data based benchmark strategy (sold line) for SEM SI.}
\end{figure*}

\begin{figure*}[htbp]
\begin{center}
	\includegraphics[scale=0.5]{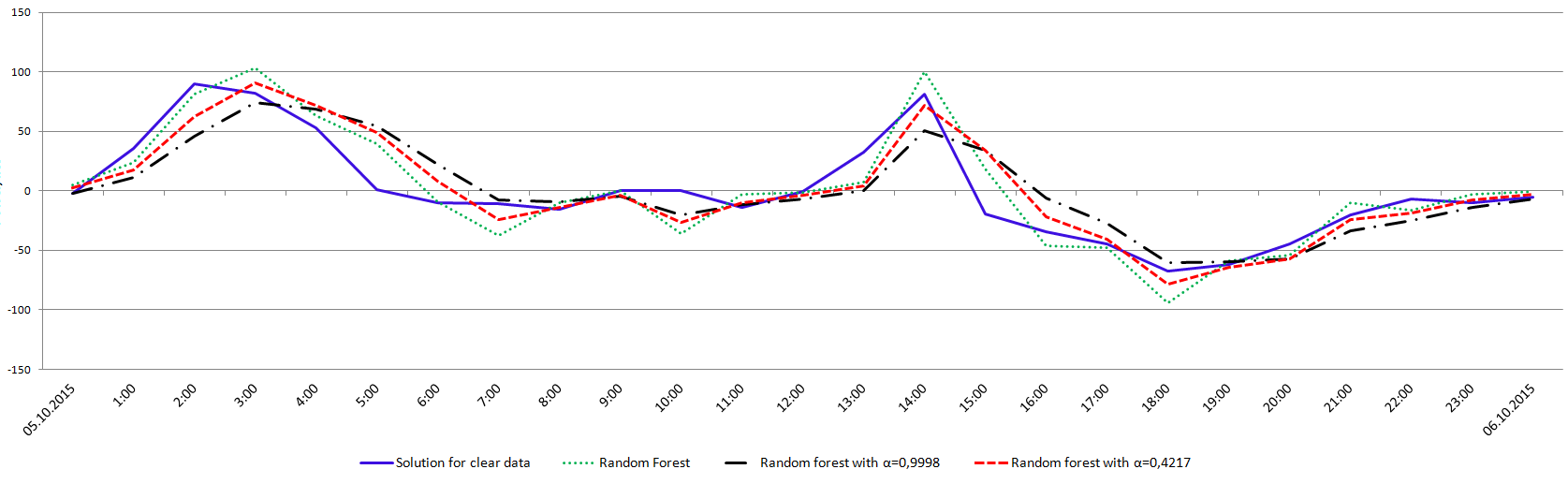}	
	\label{f5}
\end{center}
	\caption{24 hrs charge / discharge strategies using RF forecasts comparing with actual data based benchmark strategy (sold line) for SEM SI.}
\end{figure*}

%\begin{figure}[ht]
%	\includegraphics[scale=0.305]{Activity1eng.png}	
%	\label{fig:f1}
%	\caption{Activity diagram of proposed approach}
%\end{figure}

%---------------------------------------------------------------------
%----------------------  S E C T I O N 5 -----------------------------
%---------------------------------------------------------------------
\section{Conclusion}

 A novel mathematically justified approach to the load leveling problem is proposed using Volterra  models  and tested on the SEM in the Republic of Ireland. Such evolutionary models take into account both the time dependent efficiency and the availability of generation/storage of each energy storage technology in the power system. The SEM is a power system with high wind power penetration and much unpredictability due to the inherent variability of wind. The problem of efficient charge/discharge strategies is reduced to a solutions of 
linear and nonlinear integral equations and their systems. For such equations the  numerical methods are suggested finding the available storages dispatch. The proposed method is applied to real data demonstrating its efficiency. More accurate prediction can be achieved by including more representative features and handling concept drift  as suggested in \cite{pdsf}. Our future work will address UCP taking into account storages location and state of the load.

\bibliographystyle{ieeetr}
\bibliography{biblioVlad}

\begin{thebibliography}{10}

\bibitem{vse2013}
N.~I. Voropai, A.~Z. Gamm, A.~M. Glazunova, P.~V. Etingov, I.~N. Kolosok, E.~S.
  Korkina, V.~G. Kurbatsky, D.~N. Sidorov, V.~A. Spiryaev, N.~V. Tomin, R.~A.
  Zaika, and B.~Bat-Undraal, {\em Application of {M}eta-{H}euristic
  {O}ptimization {A}lgorithms in {E}lectric {P}ower {S}ystems}, pp.~564--615.
\newblock IGI Global, 2013.

\bibitem{web}
``{S}ystem {O}perator of {N}orthern {I}reland.''
  \url{http://www.soni.ltd.uk/Operations/sg/DS3/}.
\newblock Accessed: 2016-01-07.

\bibitem{bun}
D.~Bunn and E.~Farmer, ``Comparative models for electrical load forecasting,''
  {\em Int. J. Forecast}, vol.~2, pp.~501--505, 1985.

\bibitem{sor}
L.~Soares and M.~Medeiros, ``Modeling and forecasting short-term electricity
  load: a comparison of methods with a application to {B}razilian data,'' {\em
  Int. J. Forecast}, vol.~24, pp.~630--644, 2008.

\bibitem{Kalkhambkar2016}
V.~Kalkhambkar, R.~Kumar, and R.~Bhakar, ``Energy loss minimization through
  peak shaving using energy storage,'' {\em Perspectives in Science}, pp.~--,
  2016.
\newblock In press.

\bibitem{Sharma2016134}
D.~D. Sharma, S.~Singh, and J.~Lin, ``Multi-agent based distributed control of
  distributed energy storages using load data,'' {\em Journal of Energy
  Storage}, vol.~5, pp.~134 -- 145, 2016.

\bibitem{Barzin2015505}
R.~Barzin, J.~J. Chen, B.~R. Young, and M.~M. Farid, ``Peak load shifting with
  energy storage and price-based control system,'' {\em Energy}, vol.~92, Part
  3, pp.~505 -- 514, 2015.
\newblock Sustainable Development of Energy, Water and Environment Systems.

\bibitem{Gu2016723}
Y.~Gu, J.~Xu, D.~Chen, Z.~Wang, and Q.~Li, ``Overall review of peak shaving for
  coal-fired power units in china,'' {\em Renewable and Sustainable Energy
  Reviews}, vol.~54, pp.~723 -- 731, 2016.

\bibitem{Zakeri2015569}
B.~Zakeri and S.~Syri, ``Electrical energy storage systems: A comparative life
  cycle cost analysis,'' {\em Renewable and Sustainable Energy Reviews},
  vol.~42, pp.~569 -- 596, 2015.

\bibitem{Graditi2016515}
G.~Graditi, M.~Ippolito, E.~Telaretti, and G.~Zizzo, ``Technical and economical
  assessment of distributed electrochemical storages for load shifting
  applications: An italian case study,'' {\em Renewable and Sustainable Energy
  Reviews}, vol.~57, pp.~515 -- 523, 2016.

\bibitem{Parra2016130}
D.~Parra, S.~A. Norman, G.~S. Walker, and M.~Gillott, ``Optimum community
  energy storage system for demand load shifting,'' {\em Applied Energy},
  vol.~174, pp.~130 -- 143, 2016.

\bibitem{Han2015643}
X.~Han, T.~Ji, Z.~Zhao, and H.~Zhang, ``Economic evaluation of batteries
  planning in energy storage power stations for load shifting,'' {\em Renewable
  Energy}, vol.~78, pp.~643 -- 647, 2015.

\bibitem{Pazouki2016219}
S.~Pazouki and M.-R. Haghifam, ``Optimal planning and scheduling of energy hub
  in presence of wind, storage and demand response under uncertainty,'' {\em
  International Journal of Electrical Power \& Energy Systems}, vol.~80,
  pp.~219 -- 239, 2016.

\bibitem{ijai2015}
N.~Tomin, A.~Zhukov, D.~Sidorov, V.~Kurbatsky, D.~Panasetsky, and V.~Spiryaev,
  ``Random forest based model for preventing large-scale emergencies in power
  systems,'' {\em International Journal of Artificial Intelligence}, vol.~13,
  pp.~211--228, 2015.

\bibitem{mono}
D.~Sidorov, {\em Integral dynamical models: singularities, signals and
  control}, vol.~87 of {\em World Scientific Series on Nonlinear Science Series
  A}.
\newblock Singapore: World Scientific, 2015.

\bibitem{breiman2001random}
L.~Breiman, ``Random forests,'' {\em Machine learning}, vol.~45, no.~1,
  pp.~5--32, 2001.

\bibitem{friedman2001greedy}
J.~H. Friedman, ``Greedy function approximation: a gradient boosting machine,''
  {\em Annals of statistics}, pp.~1189--1232, 2001.

\bibitem{smola1997support}
A.~Smola and V.~Vapnik, ``Support vector regression machines,'' {\em Advances
  in neural information processing systems}, vol.~9, pp.~155--161, 1997.

\bibitem{StudiaInformatica}
D.~Sidorov, ``Volterra {E}quations of the {F}irst {K}ind with {D}iscontinuous
  {K}ernels in the {T}heory of {E}volving {S}ystems {C}ontrol,'' {\em Studia
  {I}nformatica {U}niversalis. {P}aris: {H}ermann {P}ubl.}, vol.~9, no.~3,
  pp.~135--146, 2011.

\bibitem{Sidorov2013}
D.~N. Sidorov, ``On parametric families of solutions of {V}olterra integral
  equations of the first kind with piecewise smooth kernel,'' {\em Differential
  Equations}, vol.~49, no.~2, pp.~210--216, 2013.

\bibitem{volt_systems}
D.~N. Sidorov, ``Solution to {S}ystems of {V}olterra {I}ntegral {E}quations of
  the {F}irst {K}ind with {P}iecewise {C}ontinuous {K}ernels,'' {\em Russian
  Mathematics (Transl. from Izvestia VUZov)}, vol.~57, no.~1, pp.~62--72, 2013.

\bibitem{mar_sid_ISU}
E.~V. Markova and D.~N. Sidorov, ``{V}olterra {I}ntegral {E}quation of the
  {F}irst {K}ind with {D}iscontinuous {K}ernels in the {T}heory of {E}volving
  {D}ynamical {S}ystems {M}odeling,'' {\em Izvestia Irkutskogo gos. univ.
  Matematika}, no.~2, pp.~31--45, 2012.

\bibitem{tyndasidmuft}
D.~N. Sidorov, A.~Tynda, and I.~Muftahov, ``Numerical solution of {V}olterra
  integral equations of the first kind with piecewise continuous kernel,'' {\em
  Vestnik {Y}u{U}r{G}u. {S}er. {M}at. {M}odel. {P}rog.}, vol.~7, no.~3,
  pp.~107--115, 2014.

\bibitem{Sidorov2014}
N.~A. Sidorov and D.~N. Sidorov, ``On the solvability of a class of {V}olterra
  operator equations of the first kind with piecewise continuous kernels,''
  {\em Mathematical Notes}, vol.~96, no.~5, pp.~811--826, 2014.

\bibitem{Markova2014}
E.~V. Markova and D.~N. Sidorov, ``On one integral {V}olterra model of
  developing dynamical systems,'' {\em Automation and Remote Control}, vol.~75,
  no.~3, pp.~413--421, 2014.

\bibitem{kantor73}
L.~V. Kantorovich and V.~I. Zhiyanov, ``Single-commodity dynamic model of the
  economy allowing for changes in asset structure in the presence of technical
  progress,'' {\em {D}okl. {A}kad. {N}auk {U}{S}{S}{R}}, vol.~211, no.~6,
  pp.~1280---1283, 1973.

\bibitem{solow}
R.~M. Solow, {\em Mathematical Methods in the Social Sciences}, ch.~Investment
  and Technical Progress, pp.~89--104.
\newblock Stanford, California: Stanford University Press, 1960.

\bibitem{glush}
V.~M. Glushkov, V.~V. Ivanov, and V.~M. Janenko, {\em Developing Systems
  Modeling}.
\newblock Moscow: Nauka, 1983.

\bibitem{Kythe}
P.~K. Kythy and P.~Puri, {\em Computational {M}ethods for {L}inear {I}ntegral
  {E}quations}.
\newblock Boston: Birkhauser, 2002.

\bibitem{muft_sid_ISU}
I.~R. Muftahov, D.~N. Sidorov, and N.~A. Sidorov, ``{L}avrentiev regularization
  of integral equations of the first kind in the space of continuous
  functions,'' {\em Izvestia Irkutskogo gos. univ. Matematika}, no.~15,
  pp.~62--77, 2016.

\bibitem{SidBlowUp}
D.~N. Sidorov, ``Existence and blow-up of {K}antorovich principal continuous
  solutions of nonlinear integral equations,'' {\em Differential Equations},
  vol.~50, no.~9, pp.~1217--1224, 2014.

\bibitem{Kantorovich2}
L.~V. Kantorovich and G.~P. Akilov, {\em Functional {A}nalyis}.
\newblock Pergamon, 2~ed., 1982.

\bibitem{eirgrid}
``{E}irgrid {G}roup. {S}ystem {I}nformation of {I}reland's {P}ower {S}ystem.''
  \url{http://www.eirgridgroup.com}.
\newblock Accessed: 2016-02-05.

\bibitem{pdsf}
A.~Zhukov, D.~Sidorov, and A.~Foley, ``Random forest based approach for concept
  drift handling,'' {\em arXiv: Artificial Intelligence (cs.AI)},
  vol.~1602.04435 [cs.AI], pp.~1--8, 2016.

\end{thebibliography}

% that's all folks
\end{document}